\documentclass[a4paper,
               biblatex,     % biblatex is used
               ]{jacow}

\usepackage{array, longtable, makecell}
\usepackage{dcolumn}% Align table columns on decimal point
\usepackage{float}
\makeatletter%
	\ifboolexpr{bool{xetex}}
	 {\renewcommand{\Gin@extensions}{.pdf,%
	                    .png,.jpg,.bmp,.pict,.tif,.psd,.mac,.sga,.tga,.gif,%
	                    .eps,.ps,%
	                    }}{}
\makeatother

\ifboolexpr{bool{xetex} or bool{luatex}} % test for XeTeX/LuaTeX
 {}                                      % input encoding is utf8 by default
 {\usepackage[utf8]{inputenc}}           % switch to utf8

\usepackage[USenglish]{babel}

\ifboolexpr{bool{jacowbiblatex}}%
 {%
  \addbibresource{ref.bib}
 }{}
\begin{document}

\title{Advanced cryogenic process control and monitoring for the SPIRAL2
superconducting LINAC}

\author{
A. Ghribi\thanks{adnan.ghribi@cnrs.fr}$^{1,2}$, M. Aburas$^{1,3}$, P-E. Bernaudin$^{1,3}$, P. Bonnay$^{3,4}$, F. Bonne$^{3,4}$, \\ A. Corbel$^{1,2}$, M. Di-Giacomo$^{1,3}$, F. Millet$^{3,4}$, A. Ngueguim Tsafak$^{3,4}$, A. Trudel$^{1,3}$, Q. Tura$^{1,2}$\\
$^{1}${Grand Accélérateur National d'Ions Lourds (GANIL), Caen, France}\\
$^{2}${Centre National de la Recherche Scientifique (CNRS - IN2P3), Caen, France}\\
$^{3}${Commissariat à l'Energie Atomique (CEA), Caen, France}\\
$^{4}${Univ. Grenoble Alpes, CEA, IRIG-DSBT, Grenoble, France}
}
\maketitle
%
%\begin{itemize}
%\item[Usage] AG [Adnan Ghribi] ; AT [Arnaud Trudel] ; MA [Muhammad Aburas] ; PB [Patrick Bonnay] ; QT [Quentin Tura] ; AC [Antoine Corbel] ; OE [Ouassim El-hasri] ; ANT [Auriol Ngueguim Tsafak]
%\item[colors] \textcolor{blue!30!black}{assignation}\\ \textcolor{gray}{non commencé}\\\textcolor{red!60!black}{en cours} \\ \textcolor{green!60!black}{terminé}
%\end{itemize}
%
\begin{abstract}
    SPIRAL2 is a superconducting accelerator for protons, deuterons and heavy ions delivering a maximum beam power of 200 kW at 40 MeV (for deuteron beams). 26 superconducting quarter wave cavities are operated at 4.4 K, plunged in a liquid helium bath with a drastic phase separator pressure control. Previous years have seen the development of advanced process control for cryogenics allowing to cope with high heat load dynamics thanks to an automatic heat dissipation compensation and a model based control. The latter is based on models, using the Simcryogenics library, optimized and linearised in the Programmable Logic Controllers. The SPIRAL2 operation has demonstrated that such control allows to keep the specified conditions for RF and beam operation even at levels of heat load dissipation approaching the physical limits of the cryogenic system. These developments allowed to synthesise a virtual observer of the dynamic heat load dissipated by the cavities. The present paper summarises the development of such observer based on the physical thermodynamic model and on machine learning techniques.
\end{abstract}

\section{INTRODUCTION}
SPIRAL2~\cite{orduz2022commissioning} at GANIL (Grand Accélérateur National d’Ions Lourds) is a state of the art accelerator that aims at delivering some of the highest intensities of rare isotope beams. Its driver should accelerate protons, deuterons and heavy ions at intensities up to 5 mA and a maximum beam power of 200 kW (see Table \ref{tab:beam_spec}) \cite{dubois2022radioactive}. The driver is composed of ECR (Electron Cyclotron Resonance) sources, a RFQ (Radio-Frequency Quadrupole) and a superconducting LINAC (LINear Accelerator). The latter is made of 26 QWR (Quater Wave Resonating) independently phased superconducting cavities. SPIRAL2 cavities are of two families: low beta ($\beta_0 = 0.07$) and high beta ($\beta_0 = 0.12$)~\cite{marchand2015performances}; where $\beta_0$ is the relative velocity of the accelerated particles. The low beta cavities are the first 12 cavities of the LINAC. They are made of bulk Niobium for their upper part and copper for the bottom part. Each low beta cavity is housed in a different cryostat and separated by room temperature quadrupole magnets. The cryostats, that we call cryomodules, integrate passive RF, vacuum and other components needed for the operation of the cavities. The last part of the LINAC comprises 7 cryomodules, housing 2 high beta cavities each. These cavities are made of bulk niobium. All cavities are operated at 88.0525 MHz. To be operated, the superconducting cavities are plunged in liquid helium baths with a stringent pressure control to avoid detuning. Every cryomodule has a dedicated satellite valves-box that insures liquid helium distribution and pressure/level regulations. These satellite cryostats form the cryodistribution. The latter can be decomposed in two branches: A left branch made of 12 valves boxes (for all low beta cryomodules) and a right branch made of 7 valves boxes (for all high beta cryomodules). More details on the cryoplant and the cryodistribution can be found in \cite{GHRIBI201744}.

The last years have seen the commissioning of the different parts of the accelerators leading to the beam commissioning and the current ramp up to achieve the target requirements \cite{orduz2022commissioning}. Cryogenics proved more challenging that initially expected~\cite{GHRIBI2020103126}. This pushed the developments of advanced model based control bringing immunity to high dynamic heat loads and uneven heat load distributions effects. One of the consequences was the ability to introduce model based virtual observers which opened the way to machine learning based diagnostics. This paper depicts these developments. The first part details the specific challenges of the cryogenics of the SPIRAL2 LINAC and the strategies to overcome them. The second focuses on virtual heat load observers.

\begin{table}
\caption{SPIRAL2 Beam Specifications}\label{tab:beam_spec}
%\begin{minipage}[b]{.48\textwidth}
\setlength\tabcolsep{1pt}
\begin{tabular*}{\columnwidth}{@{\extracolsep{\fill} }l*{4}{c}r}
\toprule
Particles  & $H^{+}$ & $^{3}He^{2+}$ & $D^{+}$ & ions & {[Units]}\\
\midrule
Q/A	& 1 & 3/2  & 1/2  & 1/3 & \\
Maximum current & 5  & 5  & 5  & 1 & {[mA]}\\
Maximum beam power & 165 & 180 & 200 & 45 & {[kW]}\\
\bottomrule
\end{tabular*}
%\end{minipage}
\end{table}

\section{CRYOGENIC CONTROL:\\ constraints and strategies}

The first purpose of the cryogenic system is to cool down the superconducting cavities under their transition temperature and keep them in optimal operating conditions. Apart from the cool down, which poses its own set of challenges due to Q-disease (see \cite{GHRIBI2020103126}), maintaining optimal operating conditions can be difficult. Superconducting cavities operate around a center frequency $f_0$ with an admissible bandwidth $\Delta f_0$. Micrometric shape deformations due to surface forces such as Lorentz forces or liquid helium pressure fluctuations can shift the cavities resonant frequency beyond their admissible bandwidth. The link between the liquid helium pressure variations $\delta P_{bath}$ and the frequency shift $\delta f$ can be translated in term of pressure-frequency sensitivity $S_p$ with:
\begin{equation}
    S_p = \frac{\delta f}{\delta P_{bath}}
\end{equation}
In the case of SPIRAL2, $S_p$ is comprised between -1 and -3 Hz/mbar for low beta cavities and between -5 and -7 Hz/mbar for high beta cavities. Considering a cavity phase shift $\Delta \Phi = \pi/12$ and a loaded quality factor $Q_L \sim 10^6$, this brings the admissible $\delta P_{bath}$ to $\sim$ 10 mbars for low beta cavities, and $\sim$ 3 mbars for high beta ones.

\begin{figure}
    \centering
    \includegraphics[width=0.5\textwidth]{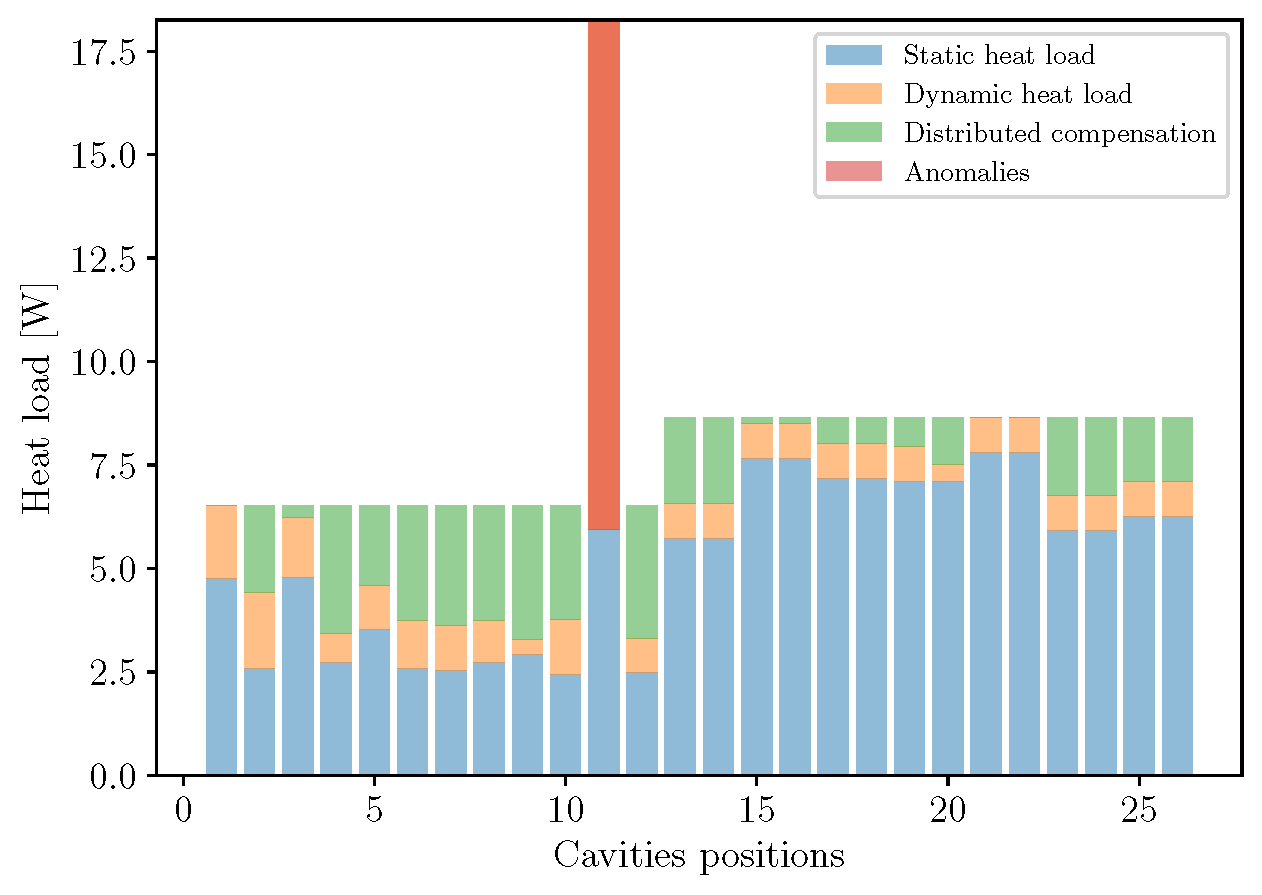}
    \caption{Heat load distribution in the LINAC showing the dynamic heat load compensation strategy.}
    \label{fig:htlc}
\end{figure}

The cryogenic system of SPIRAL2 is based on a cryoplant that produces up 120 g/s of bi-phasic helium at ~1300 mbars. The design of the system makes the pressure in the cavities helium baths (turbulence, two-phase flow behaviour) particularly difficult to control with independently tuned PID (proportional integral derivative) valves controllers and uneven helium flow distribution along the cryodistribution. To overcome these challenges, two strategies have been applied. The first one is to equally distribute the flows between the LINAC cryodistribution branches and between the cryomodules of every branch. This is done by using heaters located close to the cavities liquid helium baths to apply a so-called dynamic heat load compensation. This dynamic heat load compensation not only equally distributes the helium flows but also keeps it constant with respect to accelerating fields variations (see Fig. \ref{fig:htlc}). The second strategy was to use model based MIMO (multiple inputs - multiple outputs) controllers for simultaneous liquid helium level and pressure control. These controllers were based on a cryogenic model of the SPIRAL2 LINAC \cite{vassa2019} linearized around a given set of boundary conditions, kept constant thanks to the dynamic heat load compensation. Tuned differently, the controllers also allowed to operate abnormal cryomodules at the limit of what the physical system allows.

%\section{SYNTHESIS OF MODEL BASED CONTROL LAWS}

\section{State observers for advanced monitoring and diagnostics}
\subsection{Model based state observers}

\begin{figure}
    \centering
    \label{fig:}
    \includegraphics[width=.5\textwidth]{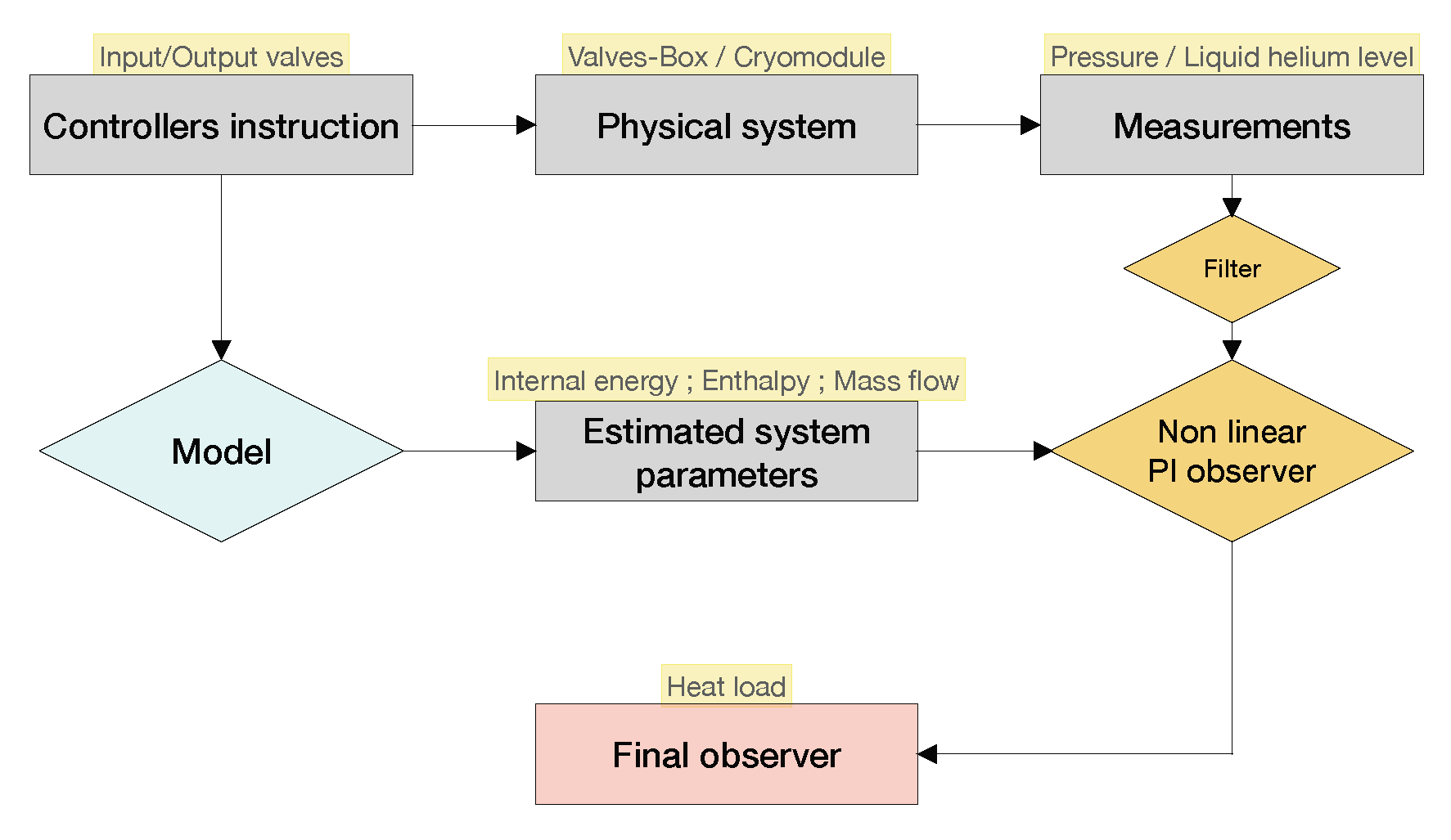}
    \caption{Model based heat load observer architecture.}\label{fig:obs_arch}
\end{figure}

\begin{figure}
	\centering
	\includegraphics[width=0.5\textwidth]{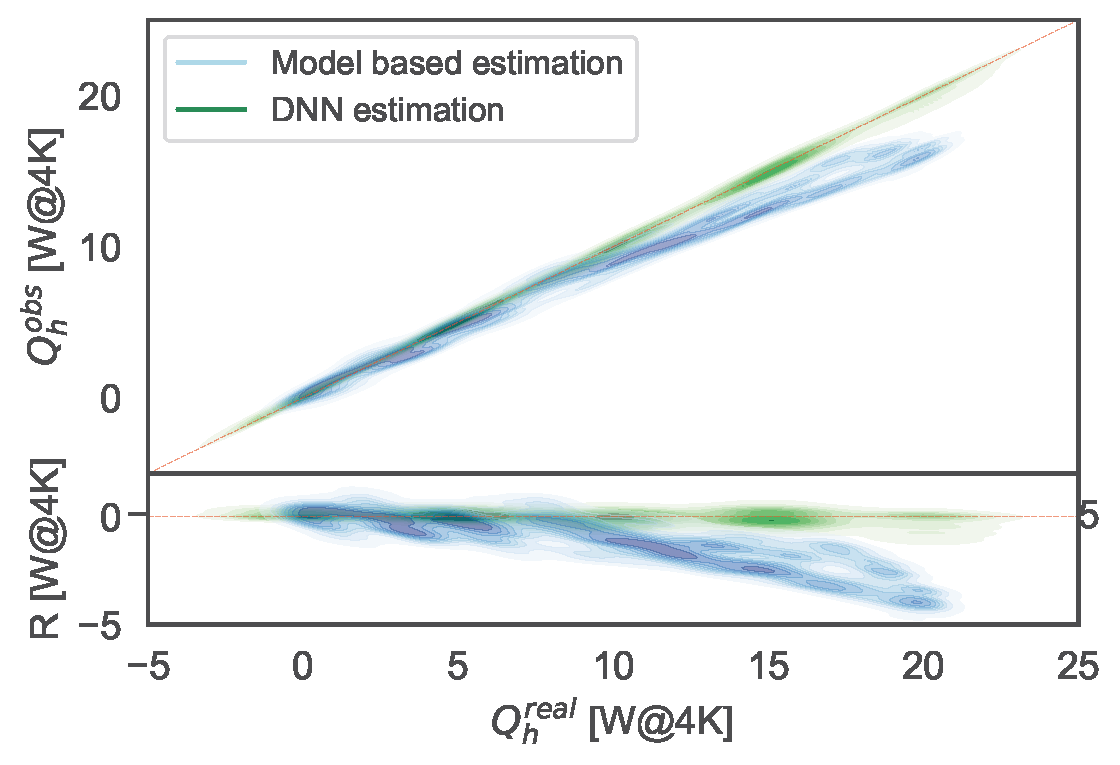}
	\caption{Up: density distribution of the estimated heat load $Q_{h}^{obs}$ vs the measured heat load $Q_{h}^{real}$. Bottom: density distribution of residual of estimated heat load. Red: ideal behaviour. Green: DNN estimator. Blue: model based linear PI estimator.}
	\label{fig:estimator}
\end{figure}

\begin{figure*}[ht]\
	\centering
	\includegraphics[width=\textwidth]{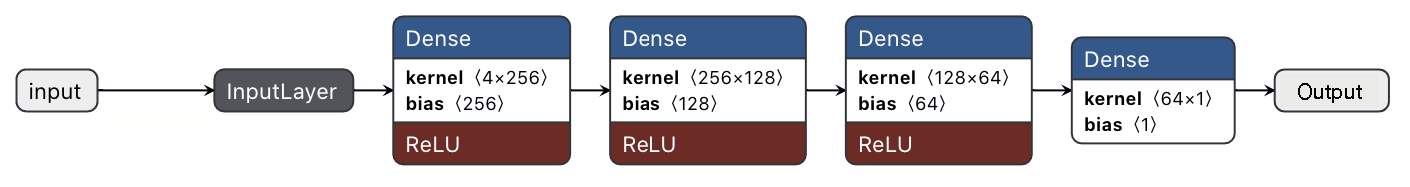}
	\caption{Neural network architecture layout used for the machine learning based heat load observer.}
	\label{fig:network}
\end{figure*}

One by-product of having a model based control is the ability to estimate hidden parameters of a physical system from existing instrumentation. In our case, the cryogenic model of the LINAC has been used to generate a virtual observer of the dynamic heat load of the cavities. The architecture of such observer is described in Fig. \ref{fig:obs_arch}. Starting from the instructions given to the input and output valves controllers (pressure and liquid helium level control), we extract the estimated internal energy of the system, its enthalpy as well as the input and output helium mass flows. Filtered pressure and liquid helium level measurements as well as estimated model parameters are injected in a PI (Proportional Integral) observer. The latter uses a time invariant linear system (obtained by linearization around the operation set-point) and is described by the following equation:
\begin{equation}
    \begin{array}{c}
        x(k+1) = Ax(k) + Bu(k) + Ed(k)\\
        y(k) = Cx(k)\\
    \end{array}
\end{equation}
where $x = \left[\begin{array}{c}
\rho\\
u_{in}
\end{array}\right]\in\mathbb{R}^{n_{x}}$ is the state vector, with $\rho$ the helium density and $u_{in}$ the internal energy; $y = \left[\begin{array}{c}
PT\\
LT
\end{array}\right]\in\mathbb{R}^{n_{y}}$ is the output vector, with $PT$ the measured helium bath pressure and $LT$ the measure liquid helium level ; $u = \left[\begin{array}{c}
\dot{m}_{in}\\
\dot{m}_{out}
\end{array}\right]\in\mathbb{R}^{n_{u}}$ the input vector, with $\dot{m}_{in}$ and $\dot{m}_{out}$ being respectively the input and output helium mass flows ; $d = \left[\begin{array}{c}
Q_{h}\\
\varepsilon_{\dot{m}}
\end{array}\right]\in\mathbb{R}^{n_{d}}$ the unknown input vector, with $Q_h$ the heat load and $\varepsilon_{\dot{m}}$ the error on the estimated helium mass flow.
Considering that the system is stable, the estimated gain of the PI observer allows to recover the unknown inputs: the heat load $Q_h$ and the mass flow error $\varepsilon_{\dot{m}}$. Results of linear PI model based heat load observers are shown in Fig. \ref{fig:estimator}. The system being non linear, a linear estimator introduces an error that increases with distance to the operation set-point. This limits its use for high dynamics fault detection. To improve the estimation of the mass flow error, a non linear PI observer is under development. Preliminary results show a clear improvement with respect to the linear PI estimator but highlight other non linearities due to measurement noise and non symmetric behaviour with regard to a positive or negative heat load dynamic.

\subsection{Machine learning state observers}

To overcome the limitations of the model based PI heat load observer, a different yet complementary approach based on machine learning techniques is being explored. A 4 layers dense neural network, following the architecture shown in Fig. \ref{fig:network}, has been used. The network was trained during 100 epochs, with a learning rate of $10^{-5}$ and a batch size of 64. A rectified linear unit (ReLu) activation function has been chosen to account for the non-linearity needed to reproduce the complex correlations underlying the features set. Finally, MSE (mean squared error) has been chosen for the estimation of the loss function with:

\begin{equation}
    \operatorname {MSE} ={\frac {1}{n}}\sum _{i=1}^{n}\left(Y_{i}-{\hat {Y_{i}}}\right)^{2}
\end{equation}

where n is the size of the dataset, $Y$ the observed values and $\hat Y$ the predicted values.

Results of the DNN model with respect to the linear PI observer can be seen in Fig. \ref{fig:estimator}. While the DNN observer outperforms the linear PI observer, it remains subject to important improvements to account for measurement noise. Features binning optimization and convolution networks are the next steps in improving such observers. The model remains important for understanding specific abnormal behaviours. In that respect, a physics informed learning is planned for precise estimation of mass flow errors.

\section{CONCLUSION}
Previous years have seen the success of the cryogenic commissioning of the SPIRAL2 LINAC. The road to operation poses a whole new set of challenges linked to the reliability of the systems. Efforts to make the commissioning a success gave birth to a robust thermodynamic model of the SPIRAL2 LINAC, allowing an efficient model based control. As a result, knowledge of the physical behaviour of the cryogenic system made possible the synthesis of virtual model based dynamic heat load observers. However, the inherent reality of everyday operation like valves mis-calibration, cold helium flow profile, instrumentation measurement noises as well as internal non linearities limit the use of a linear observer for diagnostics. A first and successful attempt to use dense neural networks for heat load virtual observers has been made. Several improvements on both machine learning and model based observers will however be necessary to enable their final integration in the accelerator main control system for fault detection and diagnostics.\\

\section{ACKNOWLEDGEMENTS}
This work has been funded by "Region Normandie" as well as the city of Caen, CNRS and CEA. This work has been done in the frame of the SPIRAL2 project and the GRAAL collaboration.

%
% only for "biblatex"
%
\ifboolexpr{bool{jacowbiblatex}}%
{\printbibliography}%
{%
	% "biblatex" is not used, go the "manual" way
	
	%\begin{thebibliography}{99}   % Use for  10-99  references
	
} % end \ifboolexpr
%
% for use as JACoW template the inclusion of the ANNEX parts have been commented out
% to generate the complete documentation please remove the "%" of the next two commands
% 
%%%\newpage

%%%\include{annexes-A4}

\end{document}